\documentclass{article}
\usepackage{spconf,amsmath,graphicx}

\usepackage{mathptmx}
\usepackage[utf8]{inputenc}
\usepackage[T1]{fontenc}
\usepackage{array}
\usepackage[margin=0.8in]{geometry}
\usepackage{tabularx}
\usepackage{graphicx}
\usepackage{caption}
\usepackage{subcaption}
\usepackage[center]{titlesec}
\usepackage{times}
\usepackage{amsmath,amssymb,amsfonts,mathrsfs}
\usepackage{float}
\usepackage{algorithm} 
\usepackage{algorithmic}
\usepackage{titlesec}

\usepackage[numbers,sort&compress]{natbib}
\usepackage{soul}
\setul{1.05mm}{}

\usepackage[english]{babel}
\usepackage[utf8]{inputenc}
\usepackage{fancyhdr}

\usepackage{amsmath} 

\usepackage{comment}
\usepackage{mathptmx}
\usepackage{graphicx}
\usepackage{paralist}
\usepackage{wrapfig}
\usepackage{color,hyperref}
\usepackage{multicol}
\usepackage{booktabs}
\usepackage{multirow}
\definecolor{darkblue}{rgb}{0.0,0.0,0.3}
\hypersetup{colorlinks,breaklinks,
            linkcolor=darkblue,urlcolor=darkblue,
            anchorcolor=darkblue,citecolor=darkblue}
            
\usepackage{url}
\usepackage{bm}
\usepackage[T1]{fontenc}
\usepackage{newtxmath,newtxtext}
\usepackage{lipsum}

\graphicspath{{figures/}}

\newcommand{\real}{\mathbb{R}}

\DeclareMathOperator{\relu}{ReLU}

\newcommand{\mcD}{\mathcal{D}}

\newcommand{\mcG}{\mathcal{G}}

\usepackage{xcolor}

\def \mbF{\mathbb{F}}
\def \mbU{\mathbb{U}}

\newcommand{\ue}{{\underline e}}

\newcommand{\ubD}{{\underline{\bm D}}}
\newcommand{\ubT}{{\underline{\bm T}}}

\newcommand{\ut}{{\underline t}}
\newcommand{\ubY}{{\underline{\bm Y}}}
\newcommand{\Rb}{{\bm R}}

\def\omg{{\Omega}}

\def \bb{\mathbf{b}}

\def \fb{\mathbf{f}}

\def \Ib{\mathbf{I}}
\def \Eb{\mathbf{E}}

\def \ub{\mathbf{u}}

\def \Ub{\mathbf{U}}

\def \Hb{\mathbf{H}}

\def \wb{\mathbf{w}}
\def \vb{\mathbf{v}}
\def \xb{\mathbf{x}}

\def \Pb{\mathbf{P}}

\def \qb{\mathbf{q}}

\def \yb{\mathbf{y}}
\def \cb{\mathbf{c}}

\def \xib{{\boldsymbol\xi}}
\def \etab{{\boldsymbol\eta}}

\newcommand{\verti}[1]{{\left\vert #1
    \right\vert}}  
    
\newcommand{\vertii}[1]{{\left\vert\left\vert #1
    \right\vert\right\vert}}

        {\begin{itemize}[#1]}{\end{itemize}%
         \vspace{-.0\baselineskip}}

        {\begin{compactitem}[#1]}{\end{compactitem}}

\usepackage[skip=0pt]{caption}

\begin{document}
\titleformat{\subsection}[hang]
   {\normalfont\bfseries\normalsize\raggedright}
   {\arabic{section}.\arabic{subsection}\centering}
   {12pt}{\normalsize }

\titleformat{\section}[hang]
   {\normalfont\bfseries\normalsize\raggedright}
   {\arabic{section}}
   {12pt}{\normalsize }

\titleformat{\subsubsection}[hang]
  {\normalfont\bfseries\normalsize\centering}
  {\arabic{section}.\arabic{subsection}.\arabic{subsubsection}}
  {12pt}{\normalsize }

\renewcommand{\figurename}{Figure}

\renewcommand{\tablename}{Table}

\renewcommand{\refname}{References}

\def\x{{\mathbf x}}
\def\L{{\cal L}}

\title{A physics-guided smoothing method for material modeling with digital image correlation (DIC) measurements}
%
 \name{Jihong Wang$^1$, Chung-Hao Lee$^2$, William Richardson$^3$, Yue Yu$^1$\thanks{Corresponding author: Yue Yu (yuy214@lehigh.edu).}
 \thanks{The authors would like to acknowledge the support from the National Institute of Health under award 1R01GM157589-01. Portions of this research were conducted on Lehigh University's Research Computing infrastructure partially supported by NSF Award 2019035.}}
 \address{$^1$Department of Mathematics, Lehigh University, Bethlehem, PA, USA\\
 $^2$Department of Bioengineering, University of California, Riverside, CA, USA\\
 $^3$Department of Chemical Engineering, University of Arkansas, Fayetteville, AR, USA}
%
%
%
%
\maketitle
\begin{abstract}
In this work, we present a novel approach to process the DIC measurements of multiple biaxial stretching protocols. In particular, we develop a optimization-based approach, which calculates the smoothed nodal displacements using a moving least-squares algorithm subject to positive strain constraints. As such, physically consistent displacement and strain fields are obtained. Then, we further deploy a data-driven workflow to heterogeneous material modeling from these physically consistent DIC measurements, by estimating a nonlocal constitutive law together with the material microstructure. To demonstrate the applicability of our approach, we apply it in learning a material model and fiber orientation field from DIC measurements of a porcine tricuspid valve anterior leaflet. Our results demonstrate that the proposed DIC data processing approach can significantly improve the accuracy of modeling biological materials.
\end{abstract}

\vspace{-5mm} 

\section{Introduction}
\label{sec:intro}

\vspace{-2.5mm}

The construction of material models from experimental spatial measurements are often needed in applications of interest to the broad scientific and engineering communities \cite{markou2021new}. These models are described as constitutive laws, in the form of a mapping from strain fields to the corresponding stress fields. In this context, digital image correlation (DIC) \cite{mccormick2010digital} is a popular optical technique to provide measurements for material displacement and strain fields, especially for highly nonlinear and heterogeneous materials. Then, the reconstructed displacement and strain fields  can be employed in physics-based \cite{tan2016numerical} or data-driven \cite{li2019predicting} approaches, to infer the underlying constitutive law and material microstructure \cite{jafarzadeh2024heterogeneous}.

However, the efficacy of this workflow may deteriorate due to unavoidable noises caused by the camera sensor and operation errors. In the former, random fluctuations are introduced in pixel intensity within an image, which greatly impact the accuracy of displacement measurements. In the latter, operators introduce human errors, such as non-even stress relaxation/heterogeneity speeds between the two testing directions, causing smooth but non-physical displacement measurements. Then, these noises/errors get amplified when calculating the derivatives and strain fields, leading to larger errors in the downstream constitutive law learning tasks. To remove the noises caused by the camera sensor, image denoising techniques have been developed \cite{fan2019brief,pan2015strain}. However, these techniques have little effects on operation errors.



In this work, we propose physics-guided smoothing (PGS) method, a novel image denoising approach to handle both artifacts in DIC measurements. In particular, we reconstruct the displacement field from DIC measurements using reproducing kernel (RK) functions as a basis \cite{you2003filters}. The RK basis coefficients are determined by enforcing the interpolation of the data and physics-based constraints. As such, the reproducing kernel serves as a low-pass filter to remove random fluctuations in pixel intensities, while the physics-based constraints help eliminate the non-physical patterns from human errors. Consequently, PGS can effectively extract displacement and strain fields from DIC measurements, inducing smaller errors in downstream constitutive law learning tasks. Our key contributions are:
\begin{itemize}
\vspace{-2.5mm}
\item We propose a novel physics-guided data processing approach for DIC measurements of biaxial stretching protocols, which we coin PGS aiming to extract smooth displacement and strain fields and eliminate non-physical features such as negative strain fields. 
\vspace{-2.5mm}
\item Our approach bridges the gap between DIC and physics-informed machine learning: i.e., PGS supplies downstream tasks with better datasets by  imposing partial physics knowledge during data-processing stage.
\vspace{-4.5mm}
\item We explore the practicality of PGS on two experimental datasets. We show that PGS effectively extracts displacement/strain fields, enabling meaningful and interpretable physical features for subsequent material constitutive law learning and microstructure discovery.
\end{itemize}

\vspace{-5mm}
\section{Background and Related Work}

\vspace{-2.5mm}

\textbf{Physics-informed machine learning: }
Learning complex physical systems from data is essential in many scientific and engineering applications \citep{ghaboussi1991knowledge,karniadakis2021physics}. Among popular approaches that use neural networks (NNs) with the purpose of learning physics,  we mention physics-informed neural networks (PINNs) and neural operators (NOs). In PINNs \cite{raissi2019physics}, the governing law is known as a given partial differential equation (PDE), then the solution of the equation is modeled by a deep NN that is designed to minimize the equation loss. This idea was also adopted into image processing pipelines to enhance performance and interpretability \cite{yi2022physical,bian2023high}. When the governing laws are unknown, NOs are an alternative method, which learns the solution operator as a mapping between infinite-dimensional function spaces \citep{li2020neural,li2020fourier}, enabling accurate and consistent predictions of continuum physical surrogates. However, vanilla NOs cannot provide interpretability of the underlying physics. 

\textbf{Constitutive operator learning: }
In order to provide physical interpretability for systems with unknown governing laws, researchers propose to learn constitutive laws \cite{tacc2023benchmarking,vlassis2020geometric,fuhg2022learning}. In the context of material modeling, that means that one should learn a mapping from the solution  (i.e., displacement/strain fields) to the corresponding loading (i.e., force/stress fields. As such, the model can be used to provide predictions of material deformation under a broad range of loading and boundary conditions. In \cite{jafarzadeh2024heterogeneous}, a constitutive operator is constructed from full-field displacement measurements. Compared to synthetic datasets, larger errors were observed when learning from DIC images due to noises and non-physical patterns.

\textbf{Image denoising and strain reconstruction techniques: }Due to unavoidable noises, the displacement gradients obtained from DIC are often seriously corrupted. To restore reasonable displacement and strain information, research efforts have been dedicated to either pre-processing the image \cite{boyat2015review,fan2019brief} or post-processing the constricted displacement/strain fields \cite{bruck1989digital,pan2015strain} with denoising techniques. Among the second category of approaches, we mention RK methods \cite{you2003filters}, where the kernels are constructed with a pre-chosen polynomial order of reproducing conditions, for ensuring desire accuracy on smooth measurements and providing a low-pass filter on noisy regions. \cite{you2022learning} showed that the RK smoothed dataset helps reduce the generalization error in downstream physics model learning tasks. However, current image denoising and strain reconstruction techniques can only handle noises in the form of \textit{random fluctuations}, and they have little effect on non-physical patterns due to operating errors.


\textbf{Mechanical testing using DIC measurements: }
In mechanical testing, material specimen is subjected to biaxial tensions simultaneously (no compressive force). However, the specimen may still experience compressive deformation due to two reasons. First, when the two applied tensions are very different in magnitude, noticeable compressive strain can be observed that is collectively due to material anisotropy and the Poisson contraction \cite{billiar2000biaxial,lyons2014mechanical}. Second, the specimen may also exhibit non-physical negative strains, especially at the relaxed state during cyclic biaxial tensions or in the direction where the dominant loading is applied. These negative strains induces non-physical biases in constitutive models, and should be eliminated in a material modeling pipeline.

\section{PGS: Physics-Guided DIC Data Processing}

\vspace{-2.5mm}


\textbf{Smoothed strain field reconstruction:} Although the proposed idea can be generalizable to other filtering techniques, as the base method we use the reproducing kernel as a filter and reconstruct the function of interest from discrete measurements. Consider a physical domain $\Omega$ with a set of given $N_{NP}$ discrete measurements, $\{\ub_{\text{exp}}(\xb_J)\}_{J=1}^{N_{NP}}$, the RK aims to construct an approximation of function $\ub(\xb)$ based on a set of kernel functions $\Phi_n^a(\xb)$ centered at points $\bar{\xb}_I$, $I=1,\cdots,N_{RK}$:\vspace{-2mm}
\begin{align}\label{eqn:localf}
&\ub(\xb)=\sum_{I=1}^{N_{RK}} \Phi_n^a(\xb;\xb-\bar{\xb}_I)\bar{\ub}_I, \\
&\text{ s.t. }\ub(\xb_J)= \ub_{\text{exp}}(\xb_J),\, J=1,\cdots, N_{NP},\vspace{-2mm}
\end{align}
with $\bar{\ub}_I$ being the coefficients to be determined and \vspace{-2mm}
\begin{equation}\label{eqn:Phi}
\Phi_n^a(\xb;\xb-\yb):=\Hb_n^T(\xb-\yb)\cb(\xb)\phi^a(\xb-\yb).\vspace{-2mm}
\end{equation}
Here, $\Hb_n=[\{(\xb-\yb)^{\bm{\alpha}}\}_{|\bm{\alpha}|\leq n}]$ contains monomial basis functions up to order $n$, $\cb$ is a vector of correction function coefficients, and $\phi^a$ is the window function with a support size of $a$. For each point $\xb$, $\cb$ is taken as $\cb(\xb)=(\mathbf{M}(\xb))^{-1}\Hb_n(0)$ with $\mathbf{M}(\xb):=\sum_{I=1}^{N_{RK}}\mathbf{H}_n(\xb-\bar{\xb}_I)\phi^a(\xb-\bar{\xb}_I)\mathbf{H}_n^{T}(\xb-\bar{\xb}_I)$, so as to guarantee the $n$-th order polynomial reproduction conditions:\vspace{-2mm}
\begin{equation}
\sum_{I=1}^{N_{RK}} \Phi_n^a(\xb;\xb-\bar{\xb}_I)p(\bar{\xb}_I)=p(\xb), \; \forall p\in [\{\xb^{\bm{\alpha}}\}_{|\bm{\alpha}|\leq n}].\vspace{-2mm}
\end{equation}

Note that $\phi^a(\xb-\xb_I)\equiv 0$ for $\xb_I\notin B_a(\xb)$, with $B_a(\xb)$ being the sphere centered at $\xb$ with a radius of $a$. The coefficients $\bar{u}_I$ associated with RK point $I$ are then determined by matching the reconstructed function with discrete measurements:
\begin{align}
&\{\bar{\ub}^\dag_I\}_{I=1}^{N_{RK}} = \underset{\{\bar{\ub}_I\}}{\text{argmin}} \; \text{loss}_\ub(\bar{\ub}_I),\label{eqn:opt_rkori}\\
&\text{loss}_\ub(\bar{\ub}_I):=\sum_{J=1}^{N_{NP}}\verti{\sum_{I=1}^{N_{RK}} \Phi^a_n(\xb_J;\xb_J-\bar{\xb}_I) \bar{\ub}_I-\ub_{\text{exp}}(\xb_J)}^2. \label{eqn:opt_u}
\end{align}
%
Applying the above approach, we can estimate the optimal weights $\bar{\ub}_I^\dag$ and the corresponding displacement field reconstructions from DIC measurements. 
The gradient of $\ub$ and the strain tensor $\Eb[\ub]$ can be obtained by:\vspace{-2mm}
\begin{align}
\nabla_\xb \ub(\xb)=&\sum_{I=1}^{N_{RK}} \nabla_\xb (\Phi_n^a(\xb;\xb-\bar{\xb}_I))\bar{\ub}_I,\\
\Eb[\ub](\xb)
&=\dfrac{1}{2}\left[(\Ib+(\nabla_\xb \ub(\xb))^T)(\Ib+\nabla_\xb \ub(\xb))-\Ib\right] \nonumber\\
&:=\left[\begin{matrix}
E_{11}[\ub](\xb) & E_{12}[\ub](\xb) \\
E_{21}[\ub](\xb) & E_{22}[\ub](\xb)
\end{matrix}\right],\vspace{-2mm}
\end{align}
where $E_{11}$, $E_{22}$ are the normal Green–Lagrange strain components in the $x$- and $y$-directions, respectively.

\textbf{Eliminating non-physical negative strain fields:} In biaxial mechanical testing, uniform tensions are applied along both the $x$- and $y$-directions. As a result, when the applied tensions are not very different in magnitude, it is unlikely to obtain a compressive strain field. However, due to operation errors, artifacts can be introduced into the strain field. As shown in Fig. \ref{fig:data1_strain}, such a non-physical strain pattern can dominate the reconstructed fields and deteriorate the accuracy.
%
Herein, we propose to eliminate the negative strains by adding a physical condition-based loss:
\vspace{-2mm}
{\small\begin{equation}\label{eqn:opt_E}
\text{loss}_\Eb(\bar{\ub}_I):=\sum_{J=1}^{N_{NP}}\verti{\relu\left(-E_{11}[ \mathbf{u}](\xb_J)\right)+\relu\left(-E_{22}[ \mathbf{u}](\xb_J)\right)}^2.
\end{equation}}
We notice that this loss is nonzero only when compressive strain patterns are reconstructed along either the $x-$ or $y-$ directions. Thus, we add $\text{loss}_\Eb$ via soft penalty constraints during RK coefficient optimization, leading us to the following minimization problem with a hybrid loss function:
\vspace{-2mm}\begin{align}\label{eq:opti}
\{\bar{\ub}^*_I\}_{I=1}^{N_{RK}} 
 =& \underset{\{\bar{\ub}_I\}}{\text{argmin}}\; (\text{loss}_\ub(\bar{\ub}_I) + \beta\,\text{loss}_\Eb(\bar{\ub}_I)).\vspace{-2mm}
\end{align}
Here, $\beta$ is a penalty parameter to eliminate the non-physical strain patterns. Thus, PGS is anticipated to enforce consistency of reconstructed displacement fields with experimental observations, while improving physical consistency.

\textbf{Optimization details: } Notice that \eqref{eq:opti} is a nonlinear optimization problem with respect to a size $O(N_{RK})$ vector. In practical applications, it often involves a large number of images with fine resolution. Solving such a nonlinear optimization may become prohibitively expensive. However, we note that \eqref{eq:opti} performs as a ``sample-wise'' optimization, and standard parallelization can be employed to split the workload to different GPUs. Moreover, the optimization problem with data loss only (see \eqref{eqn:opt_rkori}) has an analytical solution:\vspace{-2mm}
\begin{equation}\label{eqn:analytic_u}
\bar{\ub}^\dag_I=(\bm{\Phi}^\top\bm{\Phi})^{-1}\bm{\Phi}^\top \Ub_{\text{exp}},\vspace{-2mm}
\end{equation}
where $\bm{\Phi}:=[\Phi^a_n(\xb_J;\xb_J-\bar{\xb}_I)]$ is the matrix formed by kernel function values, and $\Ub_{\text{exp}}:=[\ub_{\text{exp}}(\xb_J)]$ is the vector from experimental measurements. Since the physical condition loss is necessary only when negative strain is observed, for each DIC image we first calculate the reconstructed displacement and strain fields using \eqref{eqn:analytic_u}. Then, we solve the minimization problem \eqref{eq:opti} only if negative strains are observed, and set $\bar{\ub}^\dag_I$ as the initial value of $\bar{\ub}_I$ when solving \eqref{eq:opti}. Such an initial guess practically also accelerates the optimization procedure.

We also note that the value of $\beta$ may depend on the choice of length scales in data. Taking a length scale $L$, $\text{loss}_\ub$ will be in scale $O(L^2)$ while $\text{loss}_\Eb$ is in $O(1)$, resulting in $\beta=O(L^2)$. To balance this gap in scales and provide a more robust choice of hyperparameter $\beta$, we employ a rescaling strategy, by multiplying $\beta$ with a factor $\text{loss}_{\ub}(\bar{\ub}^\dag_I)/\text{loss}_\Eb(\bar{\ub}^\dag_I)$ based on the analytical solution in \eqref{eqn:analytic_u}. The stopping criteria for $\text{loss}_{\ub}$ and $\text{loss}_{\Eb}$ are also scaled accordingly. The detailed implementation are summarized in Algorithm \ref{alg:pgs}.

\begin{algorithm}[h]
\caption{Algorithm for PGS.}
\label{alg:pgs}
\begin{algorithmic}[1]
\REQUIRE Measurements $\{\ub_{\text{exp}}(\xb_J)\}_{J=1}^{N_{NP}}$
\STATE Choose hyperparameters, including: 1) RK parameters: points $\{\bar{\xb}_I\}_{I=1}^{N_{RK}}$ and support size $a$; 2) optimization parameters: total epoch number $M$, un-normalized penalization parameter $\tilde{\beta}$, stopping criteria $\text{TOL}_{\ub}$ and $\text{TOL}_{\Eb}$.%
\STATE Compute reproducing kernel $\Phi^a$ based on \eqref{eqn:Phi}.
\STATE Compute $\bar{\ub}^\dag_I$ based on $\text{loss}_\ub$ only via a least squares method, using the analytical expression in \eqref{eqn:opt_E}. 
\STATE Compute the initial losses: $\beta_{\ub}=\text{loss}_{\ub}(\bar{\ub}^\dag_I)$ and $\beta_{\Eb}=\text{loss}_{\Eb}(\bar{\ub}^\dag_I)$; update the penalization parameter: $\beta=\tilde{\beta} \beta_{\Eb}/\beta_{\ub}$.
\STATE Set $\bar{\ub}^\dag_I$ as the initial value of $\bar{\ub}_{I}$.
\FOR{epoch number $ = 1$ to $M$}
\STATE Update the displacement field $\ub_{\text{PGS}}$ following \eqref{eqn:localf}.
\STATE Update $\text{loss}_{\ub} + \beta\,\text{loss}_{\Eb}$ following \eqref{eqn:opt_u} and \eqref{eqn:opt_E}.
\IF{$\text{loss}_{\Eb}(\bar{\ub}_I) \leq \text{TOL}_{\Eb}\beta_{\Eb}$ and $\text{loss}_{\ub}(\bar{\ub}_I) \leq \text{TOL}_{\ub}\, \beta_{\ub}$}
\STATE Break the loop and set $\bar{\ub}^*_I=\bar{\ub}_I$.
\ENDIF
\STATE Update $\bar{\ub}_I$ using Adam optimizer.
\ENDFOR
\STATE Calculate the displacement field $\ub_{\text{PGS}}$ based on $\bar{\ub}^*_I$.
\end{algorithmic}
\end{algorithm}


\textbf{Downstream constitutive operator learning:} With the updated displacement field $\ub_{\text{PGS}}$, we employ the peridynamic neural operator approach \cite{jafarzadeh2023peridynamic,jafarzadeh2024heterogeneous,wang2025monotone}, which learns a nonlocal constitutive law in the form:\vspace{-2mm}
\begin{equation}
\mcG[\ub](\xb):=\int_{{B_\delta(\xb)}}\ubT[\ub,\xb]\langle\yb-\xb\rangle-\ubT[\ub,\qb]\langle\xb-\yb\rangle\;d\yb,\vspace{-2mm}
\end{equation}
$$\ubT[\ub,\xb]\langle\xib\rangle:=t^{NN}(\omega^{NN}(\xb, \xib), \vartheta(\xb,t), \ue[\ub,\xb,t]\langle\xib\rangle, |\xib|).$$
The force state $\ubT$ function describing how the material responds under deformation. $t^{NN}$ and $\omega^{NN}$ are two trainable neural networks. Given $S_{tr}$ numbers of displacement/loading function pair samples $\{(\ub_{\text{PGS}}^s(\xb),\bb^s(\xb))\}_{s=1}^{S_{tr}}$ from smoothed reconstructions described above, ${\Pb}_{\text{exp}}^s$ as the measured stretching applied on each sample, parameters in $t^{NN}$ and $\omega^{NN}$ are obtained by minimizing \vspace{-2mm}
\begin{equation*}
   \text{loss} = \frac{\gamma}{S_{tr}}\sum_{s = 1}^{S_{tr}}\dfrac{\vertii{\mcG^{-1}[\bb^s]-\ub_{\text{PGS}}^s}_{L^2(\omg)}}{\vertii{\ub_{\text{PGS}}^s}_{L^2(\omg)}} + \frac{1 - \gamma}{S_{tr}}\sum_{s = 1}^{S_{tr}}\dfrac{\verti{{\Pb_{\exp}^s}-\Pb^s}}{\Bar{\Pb}}.\vspace{-2mm}
\end{equation*}
Here, $\gamma$ is a hyper parameter,  $\Pb^s$ denotes the spatial average of axial first Piola--Kirchhoff stresses $\Pb^s(\xb)=\int_{{B_\delta(\xb)}}\ubT[\ub,\xb]\langle\yb-\xb\rangle  \otimes (\yb-\xb) d\yb$ for sample $s$, and $\Bar{\Pb}$ is the mean of axial ground truth stresses across all training samples. Further details of the whole DIC-based constitutive operator learning pipeline can be found in Appendix Section \ref{app:PD}.

The most notable advantages of constitutive  neural operators are their resolution independence, generalizability to different input instances, and physical interpretability. In anisotropic materials like biological tissues, the strongest direction of the kernel function $\omega^{NN}$ indicates the collagen fiber orientation, and the model can discover a heterogeneous field characterizing the fiber orientation, as will be considered in our empirical studies in Section \ref{sec:exp}.

\vspace{-5mm}

\section{Empirical Studies and Discussions}\label{sec:exp}

\vspace{-2.5mm}

We assess the performance of PGS on two DIC datasets. In both datasets, the original DIC measurements are provided on a $21\times 21$ uniform mesh grid. The RK grid consists of $10\times 10$ uniform points, with the grid spacing denoted as $\Delta x$. Without otherwise stated, we sample from reconstructed displacement fields also on a $21\times 21$ original mesh grid, and supple it into the downstream constitutive operator learning tasks. In all downstream tasks, we randomly select 100 function pairs for training, 20 for validation, and use all remaining as test. The PGS method is compared with two baselines: the original dataset extracted from DIC measurement directly (referred to as ``rough''), and an RK smoothed dataset (denoted as ``smooth'') which corresponds to solving \eqref{eq:opti} without physical condition-base loss.

Our evaluation focuses on three key aspects. First, we evaluate the capability of PGS in eliminating negative strains with minimal changes on other regions. Second, we showcase the merits of physically consistent datasets in downstream constitutive operator learning tasks, by evaluating the relative $L^2$ error of displacement $\ub$ from the learnt operator. Third, we demonstrate the physical interpretability and generalization capability of learnt constitutive laws. For the former, we show the learnt kernel and discovered microstructure. For the later, we test the model on different loading scenarios and different resolutions. All experiments are conducted on a single NVIDIA GeForce RTX 3090 GPU with 24 GB memory. More details are provided in Appendix Section \ref{app:exp}.

\vspace{-5mm}

\subsection{Dataset 1: isotropic and homogeneous material}

\vspace{-2.5mm}

 The first dataset is collected from a biaxial mechanical testing on a standard nitrile glove (Dealmed, NY, USA), with total sample number $S_{tr}=600$. 

{\bf Ablation study: } 
First, we investigate PGS method's robustness across different RK smoothing levels, by choosing a support size of $a = 2.1\Delta x$, $3.1\Delta x$, and $4.1\Delta x$, respectively. As such, the reconstructed displacement field is anticipated to become smoother with the increase of $a$. In Table \ref{tab:data1_a} we present the test errors in downstream constitutive operator learning tasks. Evidently, regardless of the specific smoothing levels employed, the PGS method consistently outperforms the baseline RK smoothed method. Second, we investigate the effect of choosing different $\beta$ in Eq. \eqref{eq:opti}. A larger $\beta$ places greater emphasis on enforcing positive strain but may increase the error in $u$, causing the processed $u$ to deviate from the experimental values. Conversely, if $\beta$ is too small, the physical correction may be insufficient, leading to less physically meaningful data. This trade-off can be clearly observed in Table \ref{tab:data1_beta} and Fig. \ref{fig:data1_strain_beta} and \ref{fig:data1_u_strain_beta_1_201} (see Appendix Section \ref{app:result}).

\vspace{-2.5mm}
\begin{table}[h]
 \caption{Dataset 1: Comparison of downstream accuracy with different smoothing levels.}
    \label{tab:data1_a}
    \centering
    \begin{tabular}{|c|c|c|c|c|}
        \hline
        \multicolumn{2}{|c|}{Data type} & Train & Valid & Test \\
        \hline
        \multirow{2}{*}{$a=2.1\Delta x$} & Smooth & 9.76\% & 9.14\% & 9.13\% \\
        & PGS & 6.84\% & 7.18\% & 6.97\% \\
        \hline
        \multirow{2}{*}{$a=3.1\Delta x$} & Smooth & 9.73\% & 9.20\% & 9.14\% \\
        & PGS & 7.14\% & 7.52\% & 7.35\% \\
        \hline
        \multirow{2}{*}{$a=4.1\Delta x$} & Smooth & 9.14\% & 8.64\% & 8.60\% \\
        & PGS & 6.38\% & 6.86\% & {\bf 6.61\%} \\
        \hline
    \end{tabular}
\end{table}

\vspace{-2.5mm}
\begin{table}[h]
\caption{Dataset 1: Comparison of constitutive operator accuracy with different optimization weight $\beta$.}
    \label{tab:data1_beta}
    \centering
    \begin{tabular}{|c|c|c|c|}
        \hline
        Data type & Train & Valid & Test \\
        \hline
        Smooth($\tilde{\beta}=0$) & 9.73\% & 9.20\% & 9.14\% \\
        PGS:$\tilde{\beta}=1$ & 9.45\% & 8.99\% & 8.88\% \\
        PGS:$\tilde{\beta}=100$ & 7.14\% & 7.52\% & {\bf 7.35\%} \\
        PGS:$\tilde{\beta}=1000$ & 8.44\% & 8.73\% & 8.46\% \\
        \hline
    \end{tabular}
\end{table}

\begin{figure}[h]
    \centering
    \includegraphics[width=1.0\linewidth]{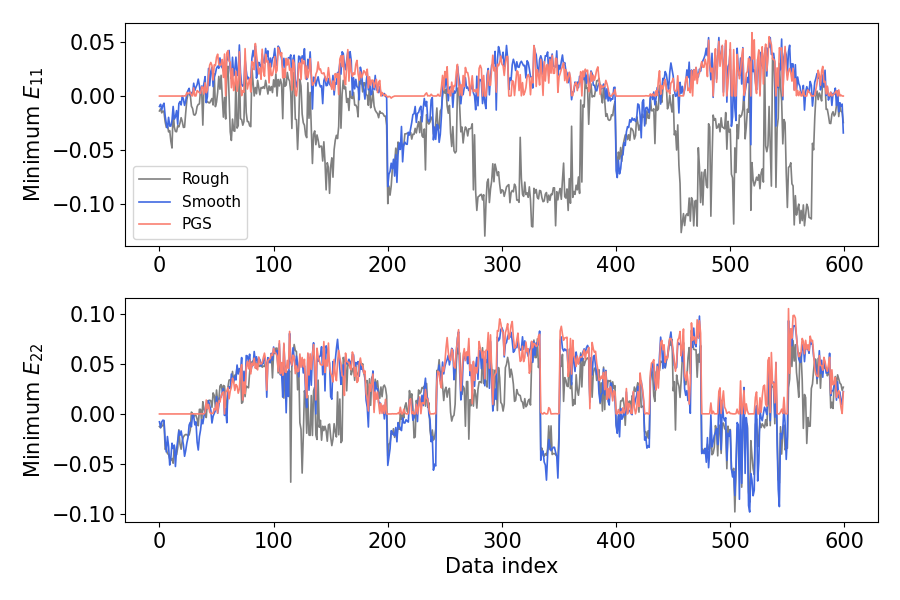}
    \caption{Dataset 1: Comparison of reconstructed strain fields.\vspace{-2.5mm}}
    \label{fig:data1_strain} 
\end{figure}

\begin{figure}[h]
    \centering
    \includegraphics[width=0.8\linewidth]{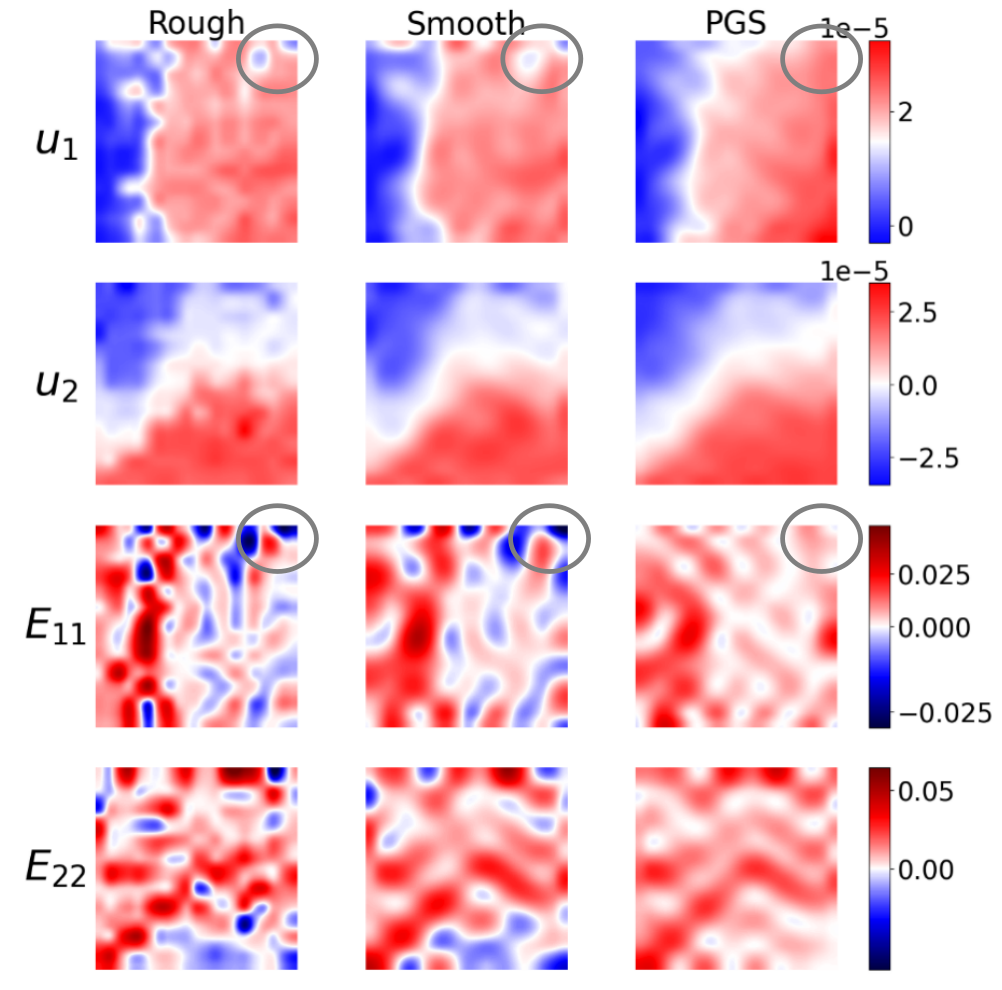}
        \caption{Dataset 1:  Comparison of displacements and strains from three types of datasets. Circles highlight a non-physical region where the negative strain needs to be eliminated. \vspace{-2.5mm}}
    \label{fig:data1_sample} 
\end{figure}

\textbf{Results:}  In the following, we choose $a=3.1\Delta x$ and $\beta=100$, and compare the performance of PGS with baselines. Fig. \ref{fig:data1_strain} presents the minimum of principal strains for each data, showing that the PGS data exhibit predominantly positive strains. Fig. \ref{fig:data1_sample} illustrates the displacements and principal strains for a representative data sample. One can see that the PGS method has successfully eliminated negative strains while smoothing the displacement fields. To further evaluate the quality of our PGS dataset, we learn the constitutive relationship using this dataset, and show results in Table \ref{tab:data1_error}: while the RK smoothing technique helps to improve test accuracy by 8.7\%, applying physics constraint further leverages the accuracy by another 20\%.


%
\begin{table}[h]
\setlength{\tabcolsep}{3pt}
    \centering
    \vspace{-2.5mm}
    \caption{Dataset 1: Comparison of the relative $L^2$ error of displacements using three types of datasets.}
    \label{tab:data1_error}
    \begin{tabular}{|c|c|c|c|c|c|}
        \hline
        Data type & Train & Valid & Test(21) & Test(16) & Test(31) \\
        \hline
        Rough & 11.01\% & 9.89\% & 10.01\% & 14.60\% & 11.06\%\\
        Smooth & 9.73\% & 9.20\% & 9.14\% & 13.54\% & 13.83\%\\
        PGS & 7.14\% & 7.52\% & {\bf 7.35\%} & {\bf 7.97\%} & {\bf 8.18\%} \\
        \hline
    \end{tabular}
\end{table}

\begin{figure}[h]
    \centering
\includegraphics[width=0.4\linewidth]{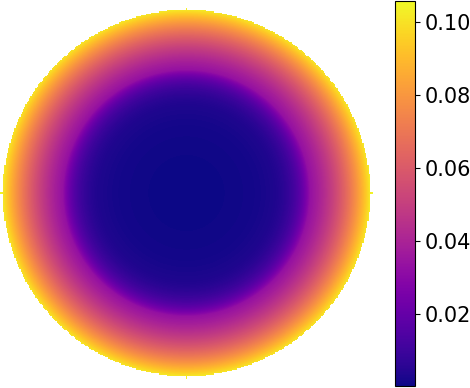}
    \includegraphics[width=0.58\linewidth]{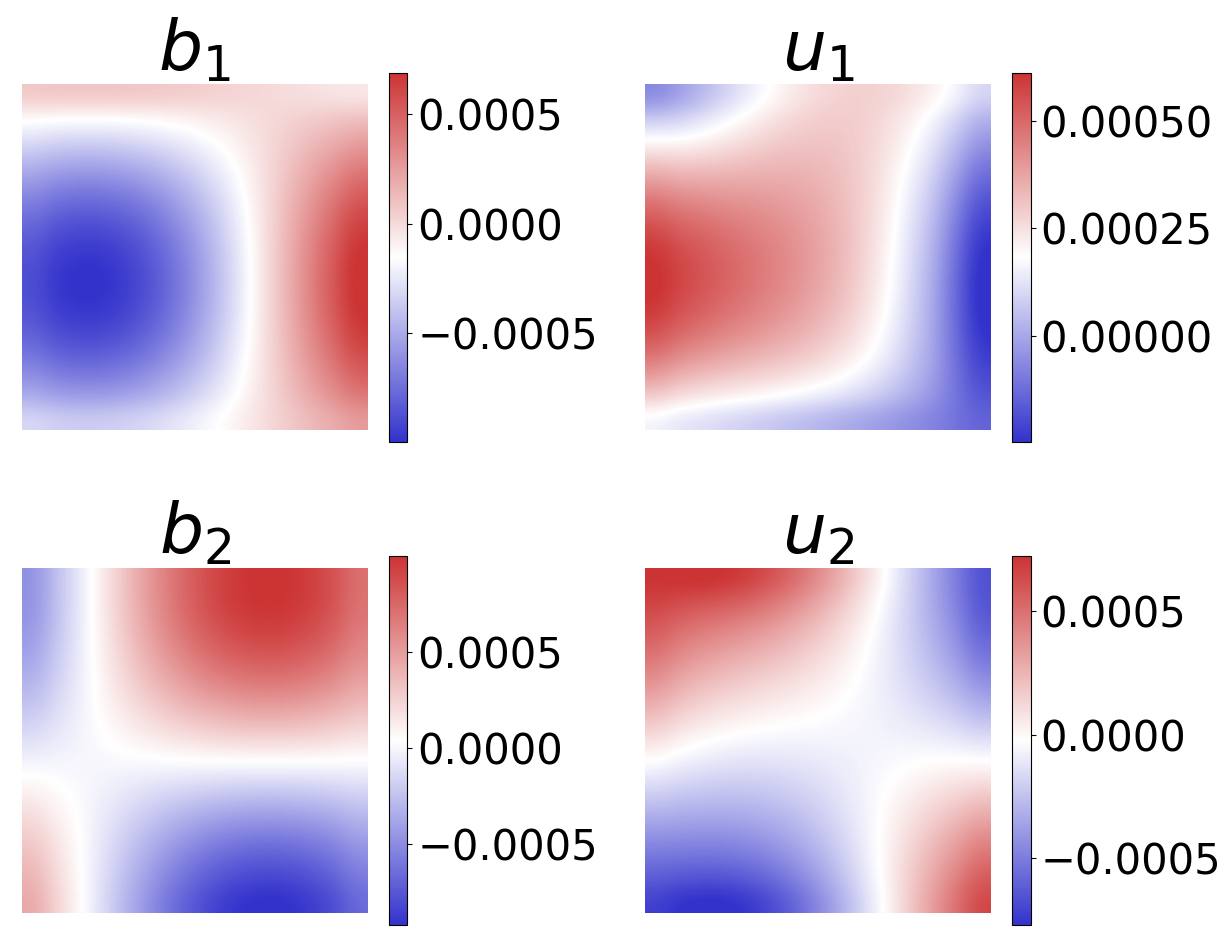}
    \caption{Dataset 1: Learnt kernel function showing the isotropic feature in this specimen (left); and the predicted displacement field $\ub$ when given a substantially different body loading $\bb$ (right), showing generalizability.\vspace{-2.5mm}}
    \label{fig:data1_kernel} 
\end{figure}

%
In Fig. \ref{fig:data1_kernel}, to demonstrate the physical interpretability of the learned model, we plot the learned kernel function. This function characterizes the interaction between neighboring points, showing the isotropic property of this specimen. Additionally, once the model is trained, it can be used to solve for the displacement given a new and unseen loading $\bb$. Herein, the training data only includes zero loading ($\bb=0$) samples. To verify this capability, we choose a non-zero loading field $\bb(\xb) = [\cos(x_1)\cos(x_2), 0]$ and solve for the displacement field using the learnt constitutive model. The loading $\bb$ and the solved displacement $\ub$ are plotted. Last but not least, we also investigate the generalizability of learnt model to data with different resolutions. We train the model on $21\times 21$ grids, then test it on coarser ($16\times 16$) and finer ($31\times 31$) grids. As shown in Table \ref{tab:data1_error}, the model generated from PGS datasets is the least sensitive to the change of data resolutions.

\vspace{-5mm}

\subsection{Dataset 2: anisotropic and heterogeneous material}

\vspace{-2.5mm}

We now consider a biaxial mechanical testing data from a representative porcine tricuspid valve anterior leaflet (TVAL) tissue. The data collection procedure can be found in Appendix Section \ref{app:data}.
Seven displacement-controlled biaxial tension protocols were performed, with the following biaxial tension ratios: $F_{x}:F_{y}=1:1,1:0.75,1:0.5,1:0.25,0.75:1,0.5:1,$ and $0.25:1$. Consequently, negative $E_{11}$ values are non-physical for the first four loading protocols, while negative $E_{22}$ values are non-physical for the last three loading protocols.
Therefore, we only need to include $E_{11}$ in loss strain \eqref{eqn:opt_E} for the first four loadings, and $E_{22}$ for the last three loadings. The total sample size is $S_{tr}=1926$.

\begin{figure}[h]
    \centering
    \includegraphics[width=1.0\linewidth]{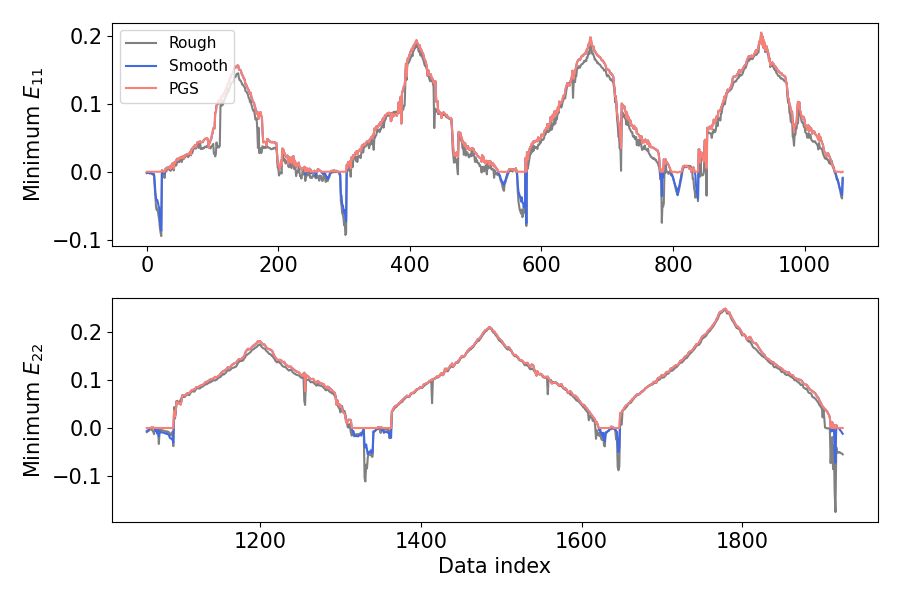}
    \caption{Dataset 2: Comparison of the minimum strain from three datasets.\vspace{-2.5mm}}
    \label{fig:data2_strain} 
\end{figure}

\textbf{Results:} Fig. \ref{fig:data2_strain} illustrates the minimum strain values of the processed data. The PGS method again ensures that the strain field is almost entirely positive, as desired. The test errors of the constitutive operator model trained using PGS and two baselines are summarized in Table \ref{tab:data2_error}: PGS again achieves the best test performance. However, it should be noted that in this dataset, as can be observed from Fig. \ref{fig:data2_strain}, there are a few non-physical data points. Therefore, the training error does not show a significant improvement compared with the two baselines. 
Moreover, to evaluate the performance of the trained models on original non-physical data, we design a special dataset that includes all data points with minimum strain $E_{11}<0$ for the first four loadings and strain $E_{22}<0$ for the last three loadings. This dataset, which contains a total of 332 data samples, is denoted as ``STest''. As can be seen from Table \ref{tab:data2_error}, for this challenging dataset, the model trained with PGS data demonstrates an improved accuracy.

\begin{figure}[h]
    \centering
    \includegraphics[width=0.48\linewidth]{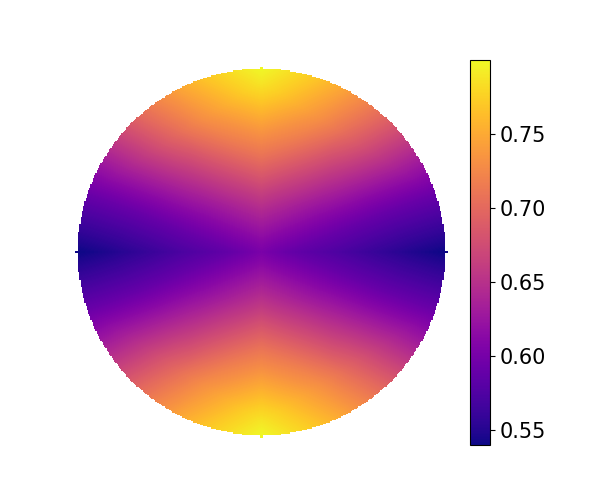}
    \includegraphics[width=0.48\linewidth]{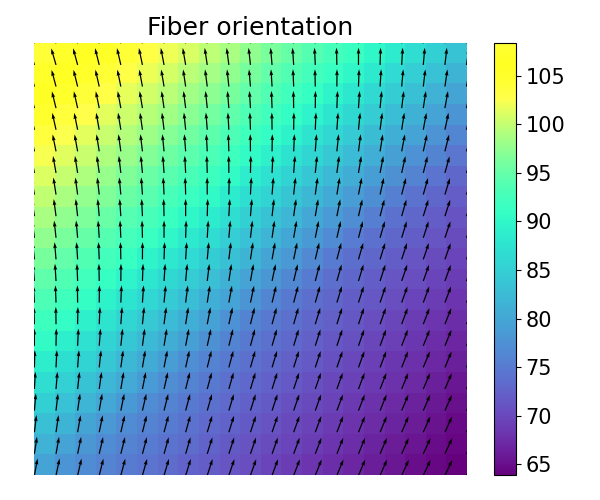}
    \caption{Dataset 2: Learnt kernel function and fiber orientation, showing the anisotropic and heterogeneous nature of this material.\vspace{-2.5mm}}
    \label{fig:data2_kernel_alpha} 
\end{figure}

Additionally, we visualize the learned kernel function together the inferred fiber orientations at each position in Fig. \ref{fig:data2_kernel_alpha}, providing physical interpretability about the anisotropic and heterogeneous nature of biological tissues.

\begin{table}[h]
\caption{Dataset 2: Comparison of the relative $L^2$ error of displacements using three types of datasets.}
    \label{tab:data2_error}
    \centering
    \begin{tabular}{|c|c|c|c|c|}
        \hline
        Data type & Train & Valid & Test & STest(332) \\
        \hline
        Rough & 6.85\% & 8.71\% & 6.21\% & 17.85\% \\
        Smooth & 5.62\% & 7.61\% & 5.13\% & 16.85\% \\
        PGS & 5.49\% & 7.33\% & {\bf 5.11\%}  & {\bf 14.68\%}\\
        \hline
    \end{tabular}
\end{table}
\vspace{-2.5mm}

\vspace{-5mm}

\section{Conclusion}

\vspace{-2.5mm}

We present physics-guided smoothing (PGS) method, a novel mechanical testing data processing approach for DIC measurements. PGS leverages kernel-based displacement and strain reconstruction techniques with partial physics knowledge, so as to eliminates both the random noises from sensors and the smooth non-physical strain patterns caused by operation errors. Then, the reconstructed displacement fields feature better physical consistency and improve the learning accuracy in downstream tasks, e.g., for material constitutive law and microstructure learning. We perform empirical evaluations on two real-world DIC datasets, showing the improved physical consistency and generalizability.




\bibliographystyle{IEEEbib}
\bibliography{yyu}

\begin{thebibliography}{10}

\bibitem{markou2021new}
George Markou, Reyes Garcia, Christos Mourlas, Maurizio Guadagnini, Kypros
  Pilakoutas, and Manolis Papadrakakis,
\newblock ``A new damage factor for seismic assessment of deficient bare and
  frp-retrofitted rc structures,''
\newblock {\em Engineering Structures}, vol. 248, pp. 113152, 2021.

\bibitem{mccormick2010digital}
Nick McCormick and Jerry Lord,
\newblock ``Digital image correlation,''
\newblock {\em Materials today}, vol. 13, no. 12, pp. 52--54, 2010.

\bibitem{tan2016numerical}
Xin Tan, Heinz Konietzky, and Wei Chen,
\newblock ``Numerical simulation of heterogeneous rock using discrete element
  model based on digital image processing,''
\newblock {\em Rock Mechanics and Rock Engineering}, vol. 49, pp. 4957--4964,
  2016.

\bibitem{li2019predicting}
Xiang Li, Zhanli Liu, Shaoqing Cui, Chengcheng Luo, Chenfeng Li, and Zhuo
  Zhuang,
\newblock ``Predicting the effective mechanical property of heterogeneous
  materials by image based modeling and deep learning,''
\newblock {\em Computer Methods in Applied Mechanics and Engineering}, vol.
  347, pp. 735--753, 2019.

\bibitem{jafarzadeh2024heterogeneous}
Siavash Jafarzadeh, Stewart Silling, Lu~Zhang, Colton Ross, Chung-Hao Lee,
  SM~Rahman, Shuodao Wang, and Yue Yu,
\newblock ``Heterogeneous peridynamic neural operators: Discover biotissue
  constitutive law and microstructure from digital image correlation
  measurements,''
\newblock {\em arXiv preprint arXiv:2403.18597}, 2024.

\bibitem{fan2019brief}
Linwei Fan, Fan Zhang, Hui Fan, and Caiming Zhang,
\newblock ``Brief review of image denoising techniques,''
\newblock {\em Visual Computing for Industry, Biomedicine, and Art}, vol. 2,
  no. 1, pp. 7, 2019.

\bibitem{pan2015strain}
Bing Pan, Jieyu Yuan, and Yong Xia,
\newblock ``Strain field denoising for digital image correlation using a
  regularized cost-function,''
\newblock {\em Optics and Lasers in Engineering}, vol. 65, pp. 9--17, 2015.

\bibitem{you2003filters}
Yang You, J-S Chen, and Hongsheng Lu,
\newblock ``Filters, reproducing kernel, and adaptive meshfree method,''
\newblock {\em Computational mechanics}, vol. 31, no. 3, pp. 316--326, 2003.

\bibitem{ghaboussi1991knowledge}
J~Ghaboussi, JH~Garrett~Jr, and Xiping Wu,
\newblock ``Knowledge-based modeling of material behavior with neural
  networks,''
\newblock {\em Journal of engineering mechanics}, vol. 117, no. 1, pp.
  132--153, 1991.

\bibitem{karniadakis2021physics}
George~Em Karniadakis, Ioannis~G Kevrekidis, Lu~Lu, Paris Perdikaris, Sifan
  Wang, and Liu Yang,
\newblock ``Physics-informed machine learning,''
\newblock {\em Nature Reviews Physics}, vol. 3, no. 6, pp. 422--440, 2021.

\bibitem{raissi2019physics}
Maziar Raissi, Paris Perdikaris, and George~E Karniadakis,
\newblock ``Physics-informed neural networks: {A} deep learning framework for
  solving forward and inverse problems involving nonlinear partial differential
  equations,''
\newblock {\em Journal of Computational Physics}, vol. 378, pp. 686--707, 2019.

\bibitem{yi2022physical}
Xinyu Yi, Yuxiao Zhou, Marc Habermann, Soshi Shimada, Vladislav Golyanik,
  Christian Theobalt, and Feng Xu,
\newblock ``Physical inertial poser (pip): Physics-aware real-time human motion
  tracking from sparse inertial sensors,''
\newblock in {\em Proceedings of the IEEE/CVF conference on computer vision and
  pattern recognition}, 2022, pp. 13167--13178.

\bibitem{bian2023high}
Liheng Bian, Haoze Song, Lintao Peng, Xuyang Chang, Xi~Yang, Roarke Horstmeyer,
  Lin Ye, Chunli Zhu, Tong Qin, Dezhi Zheng, et~al.,
\newblock ``High-resolution single-photon imaging with physics-informed deep
  learning,''
\newblock {\em Nature Communications}, vol. 14, no. 1, pp. 5902, 2023.

\bibitem{li2020neural}
Zongyi Li, Nikola Kovachki, Kamyar Azizzadenesheli, Burigede Liu, Kaushik
  Bhattacharya, Andrew Stuart, and Anima Anandkumar,
\newblock ``Neural operator: Graph kernel network for partial differential
  equations,''
\newblock {\em arXiv preprint arXiv:2003.03485}, 2020.

\bibitem{li2020fourier}
Zongyi Li, Nikola~Borislavov Kovachki, Kamyar Azizzadenesheli, Kaushik
  Bhattacharya, Andrew Stuart, and Anima Anandkumar,
\newblock ``Fourier {N}eural{O}perator for {P}arametric {P}artial
  {D}ifferential {E}quations,''
\newblock in {\em International Conference on Learning Representations}, 2020.

\bibitem{tacc2023benchmarking}
Vahidullah Ta{\c{c}}, Kevin Linka, Francisco Sahli-Costabal, Ellen Kuhl, and
  Adrian~Buganza Tepole,
\newblock ``Benchmarking physics-informed frameworks for data-driven
  hyperelasticity,''
\newblock {\em Computational Mechanics}, pp. 1--17, 2023.

\bibitem{vlassis2020geometric}
Nikolaos~N Vlassis, Ran Ma, and WaiChing Sun,
\newblock ``Geometric deep learning for computational mechanics part i:
  Anisotropic hyperelasticity,''
\newblock {\em Computer Methods in Applied Mechanics and Engineering}, vol.
  371, pp. 113299, 2020.

\bibitem{fuhg2022learning}
Jan~N Fuhg, Nikolaos Bouklas, and Reese~E Jones,
\newblock ``Learning hyperelastic anisotropy from data via a tensor basis
  neural network,''
\newblock {\em Journal of the Mechanics and Physics of Solids}, vol. 168, pp.
  105022, 2022.

\bibitem{boyat2015review}
Ajay~Kumar Boyat and Brijendra~Kumar Joshi,
\newblock ``A review paper: noise models in digital image processing,''
\newblock {\em arXiv preprint arXiv:1505.03489}, 2015.

\bibitem{bruck1989digital}
HA~Bruck, SR~McNeill, Michael~A Sutton, and WH~Peters,
\newblock ``Digital image correlation using newton-raphson method of partial
  differential correction,''
\newblock {\em Experimental mechanics}, vol. 29, pp. 261--267, 1989.

\bibitem{you2022learning}
Huaiqian You, Quinn Zhang, Colton~J Ross, Chung-Hao Lee, and Yue Yu,
\newblock ``Learning deep implicit fourier neural operators ({IFNOs}) with
  applications to heterogeneous material modeling,''
\newblock {\em Computer Methods in Applied Mechanics and Engineering}, vol.
  398, pp. 115296, 2022.

\bibitem{billiar2000biaxial}
Kristen~L Billiar and Michael~S Sacks,
\newblock ``Biaxial mechanical properties of the natural and glutaraldehyde
  treated aortic valve cusp—part i: experimental results,''
\newblock {\em J. Biomech. Eng.}, vol. 122, no. 1, pp. 23--30, 2000.

\bibitem{lyons2014mechanical}
Mathew Lyons, Des~C Winter, and Ciaran~K Simms,
\newblock ``Mechanical characterisation of porcine rectus sheath under uniaxial
  and biaxial tension,''
\newblock {\em Journal of biomechanics}, vol. 47, no. 8, pp. 1876--1884, 2014.

\bibitem{jafarzadeh2023peridynamic}
Siavash Jafarzadeh, Stewart Silling, Ning Liu, Zhongqiang Zhang, and Yue Yu,
\newblock ``Peridynamic neural operators: A data-driven nonlocal constitutive
  model for complex material responses,''
\newblock {\em Computer Methods in Applied Mechanics and Engineering}, vol.
  425, pp. 116914, 2024.

\bibitem{wang2025monotone}
Jihong Wang, Xiaochuan Tian, Zhongqiang Zhang, Stewart Silling, Siavash
  Jafarzadeh, and Yue Yu,
\newblock ``Monotone peridynamic neural operator for nonlinear material
  modeling with conditionally unique solutions,''
\newblock {\em arXiv preprint arXiv:2505.01060}, 2025.

\end{thebibliography}

\appendix
\onecolumn 


\section*{}
\vspace{1cm}
\begin{center}
    \Large \textbf{Appendix of “A physics-guided smoothing method for material modeling with digital image correlation measurements”}
\end{center}

\section{Constitutive Law Learning} \label{app:PD}

Instead of learning a PDE solution operator which maps from a body load function $\bb$ to the desired solution (displacement) field $\ub$, constitutive operator learning approaches \cite{jafarzadeh2024heterogeneous} aim to construct a surrogate operator $\mcG:\mbU\rightarrow \mbF$ that maps from the displacement function $\ub(\xb)$ to the body load function $\bb(\xb)$. Without loss of generality, in this work we consider the constitutive operator as a peridynamic material model. For simplicity, we focus on (quasi)static and two-dimensional tasks.

\subsection{Peridynamic Theory}

Peridynamics is a continuum mechanics model, where the constitutive operator is written as an integral instead of differential operators as in classical PDEs. A quasi-static peridynamics model writes:
\begin{equation}
\int_{{B_\delta(\xb)}}\fb(\ub,\qb,\xb)\;d\qb+\bb(\xb) =\bm{0}\text{,}\quad \xb\in\omg\text{,}
\label{eqn-pdeomy}
\end{equation}
where $\omg\subset\real^d$ is a domain of interest, $\xb$ and $\qb$ are material points in the reference (undeformed) configuration of the body. $B_\delta(\xb)$ is a ball centered at $\xb$ of radius $\delta$.  $\ub$ is the displacement field,
$\bb(\xb,t)$ is the body force density (external loading).
$\fb(\ub,\qb,\xb)$ is the pairwise bond force density that $\qb$ exerts on $\xb$, satisfying
$\fb(\ub,\qb,\xb)=-\fb(\ub,\xb,\qb).$
The pairwise bond force density is given by
\begin{equation}
  \fb(\ub,\qb,\xb)=
  \ubT[\ub,\xb]\langle\qb-\xb\rangle-\ubT[\ub,\qb]\langle\xb-\qb\rangle\text{ ,}
\label{eqn-Tdef}
\end{equation}
where the underlined symbols denote states.
A material model \(\hat\ubT(\ubY)\) determines the force state \(\ubT\) based on the deformation state \(\ubY\), which is characterized by
\begin{equation}
   \ubY[\ub,\xb]\langle\qb-\xb\rangle=\xib+\etab, \text{ where }\xib:=\qb-\xb,\;\etab:=\ub(\qb)-\ub(\xb)\text{.}
\label{eqn-Ydef}
\end{equation}
 Denoting the unit direction of the deformed bond as:
\begin{equation}
   \ubD[\ub,\xb]\langle\qb-\xb\rangle:=\dfrac{\ubY[\ub,\xb]\langle\qb-\xb\rangle}{\verti{\ubY[\ub,\xb,t]\langle\qb-\xb\rangle}}=\dfrac{\xib+\etab}{\verti{\xib+\etab}},
\label{eqn-Mdef}
\end{equation}
and the length changes of the bond as
\begin{equation}
   \ue[\ub,\xb]\langle\qb-\xb\rangle:=\verti{\xib+\etab}-\verti{\xib},
\label{eqn-edef}
\end{equation}
then the material model for a heterogeneous body composed of ordinary, mobile material can be written as:
\begin{equation}
 \hat\ubT(\ubY,\xb) = \ut(\ue,\xb)\,\ubD\text{ ,}
\label{eqn-mobile}
\end{equation}
where $\ut$ denotes the scalar force state. This formulation guarantees linear and angular momentum conservation, Galilean invariance, and frame invariance (objectivity).

Then we obtain the peridynamic model: 
\begin{equation}
\mcG[\ub](\xb)+\bb(\xb) = \bm{0}\text{, for } \xb\in\Omega,\label{eqn:peri_full}
\end{equation}
where the operator $\mcG$ is formulated as:
\begin{equation}\label{eqn:G}
\mcG[\ub](\xb):=\int_{{B_\delta(\mathbf{0})}}\left(\ut[\ub,\xb]\langle\xib\rangle+\ut[\ub,\xb+\xib]\langle -\xib\rangle\right) \ubD[\ub,\xb]\langle\xib\rangle\;d\xib,
\end{equation}
with boundary conditions are supplied by
\begin{equation}
\ub(\xb)=\ub_{BC}(\xb) \text{, for }\xb\in\Omega_I,
\end{equation}
where $\Omega_I:=\{\xb|\xb\in\real^d \backslash \omg,\,\text{dist}(\xb,\omg)<2\delta\}$ is the interaction region in which boundary data $\ub_{BC}$ is prescribed. 

\subsection{Peridynamics operator learning (PNO)} \label{app:PNO}

To learn a constitutive operator based on peridynamic theory, one can parameterize the scalar force state $\ut$ with neural networks: \begin{equation}\label{eqn:umt}
  \ut[\ub,\xb]\langle{\xib}\rangle:= t^{NN}(\omega(\xb, \xib), \vartheta(\xb), \ue[\ub,\xb]\langle\xib\rangle, |\xib|;\vb)\text{ ,}
\end{equation}
where
\begin{equation}\label{eqn:omega}
  \omega(\xb, \xib) :=\omega^{NN}(\Rb(-\alpha(\xb)) \xib;\wb)\text{ ,}
\end{equation}
\begin{equation}\label{eqn:Rot_mat}
  \Rb(\theta) := 
\begin{bmatrix}
    \mathrm{cos} \theta & -\mathrm{sin} \theta \\
    \mathrm{sin} \theta & \mathrm{cos} \theta
\end{bmatrix},
\end{equation}
\begin{equation}\label{eqn:dila}
  \vartheta(\xb) :=\dfrac{\int_{B_\delta (\mathbf{0})} \omega^{NN}\left(\xib;\wb\right)\ue[\ub,\xb]\langle\xib\rangle|\xib| d \xib}{\int_{B_\delta (\mathbf{0})} \omega^{NN}\left(\xib;\wb\right)|\xib|^2 d \xib}\text{ .}
\end{equation}
Here, $t^{NN}$ and $\omega^{NN}$ are scalar-valued functions implemented as multi-layer perceptrons (MLPs) with learnable parameters $\vb$ and $\wb$, respectively.
The function $\omega$ serves as a kernel that defines the weighting of neighboring material points. To account for fiber orientation at each location, the rotation matrix $\Rb$ is introduced to capture the fiber orientation at each location by aligning the direction of the kernel function with the fiber orientation. If the material is homogeneous, then $\omega$ can only depend on the direction of the bond vector $\xib$ as well as $|\xib|$, i.e., $\omega(\xb, \xib) :=\omega(\xib)$.  Here $\vartheta$ represents a nonlocal generalization of the dilatation, which describes the volume change of material near a point due to the deformation in volume. 

With the PNO architecture, one can model complex material mechanical responses learned from data. In particular, given a set of function pair observations $\mcD=\{\ub^s(\xb),\bb^s(\xb)\}_{s=1}^{S_{tr}}$ of the loading field $\bb^s(\xb)$ and the corresponding displacement field $\ub^s(\xb)$, the set of parameters in the network architecture is inferred by minimizing the error in \eqref{eqn:peri_full}. 

When $\bb\neq 0$, one can use the relative $L^2$ error of the output function, $\bb$, as the loss function:
\begin{equation}\label{eqn: caseI_loss}
    \text{loss}_{\bb} = \frac{1}{S_{tr}}\sum_{s = 1}^{S_{tr}}\dfrac{\vertii{\mcG[\ub^s]+\bb^s}_{L^2(\omg)}}{\vertii{\bb^s}_{L^2(\omg)}}\text{ .} 
\end{equation}
However, in biaxial testing protocols we are considering here, external forces are absent. Then, the loss function \eqref{eqn: caseI_loss} becomes invalid due to a zero denominator. To overcome this issue, we reformulate the loss function using the displacement field and the averaged axial components of the first Piola–Kirchhoff stress:
\begin{equation}\label{eqn: caseII_loss}
   \text{loss} = \frac{\gamma}{S_{tr}}\sum_{s = 1}^{S_{tr}}\dfrac{\vertii{\mcG^{-1}[\bb^s]+\ub^s}_{L^2(\omg)}}{\vertii{\ub^s}_{L^2(\omg)}} + \frac{1 - \gamma}{S_{tr}}\sum_{s = 1}^{S_{tr}}\dfrac{\verti{{\Pb_{\text{exp}}^s}-\Pb^s}}{\Bar{\Pb}}\text{ ,}
\end{equation}
where ${\Pb_{\text{exp}}^s}$ denotes the spatial average of axial first Piola--Kirchhoff stresses for sample $s$, and $\Bar{\Pb}$ is the mean of axial ground truth stresses across all training samples, $\gamma$ is a tunable hyperparameter. 
Once the constitutive law is obtained, for any new loading instance $\bb(\xb)$, we solve for the displacement field $\ub(\xb)$ using an iterative nonlinear static solver. Hence, the PNO model is generalizable to new and unseen loading instances.

\section{DIC Datasets} \label{app:data}

Traditional strain gauges, working well for engineering materials (e.g., aluminum/steel), are not capable of measuring the strain of biological tissues typically under a large deformation regime. On the other hand, laser or optical extensometer are the experimental alternative to measure large-deformation strains of bio-tissues, but they are limited to 1D axial strain measurements. Since most of the soft bio-tissues are highly nonlinear and heterogeneous in nature, digital image correlation (DIC) is a more commonly adopted displacement- and strain-tracking technique that can be integrated with the planar biaxial testing system – our application focus.

\subsection{Data Collection for dataset 1}

This dataset is generated from a biaxial testing on a standard nitrile glove sourced from Dealmed (New York, USA), cutting a 7.5 mm by 7.5 mm specimen for analysis. The specimen’s thickness was precisely measured using a Keyence laser thickness gauge (Illinois, USA). We then applied a random speckle pattern to the surface using an airbrush.

The speckled specimen was mounted on a biaxial testing device from CellScale Biomaterials Testing Co. (Canada), using five BioRake tines for secure fixation. During the biaxial testing, three loading-unloading cycles were performed in each direction, with a target force of 750 mN. Force and actuator position data were recorded at 5 Hz for stress and strain calculations, essential for the constitutive model fitting.

Simultaneously, a CCD camera captured images at 5 Hz, and digital image correlation (DIC) analysis was conducted using the CellScale LabJoy software. The central 6 mm by 5.5 mm region of the specimen, which had a more uniformly distributed speckle pattern, was selected for tracking. A 20 by 20 node grid was constructed, and the tracked coordinates were exported for further analysis.


\subsection{Data Collection for dataset 2}

This dataset is from a biaxial mechanical testing on porcine tricuspid valve anterior leaflet (TVAL) tissue. The tissue was sourced from an adult porcine heart (120 kg, 1.5 years old) obtained from a USDA-approved abattoir. Upon arrival at the laboratory, the TVAL tissue was sectioned into square specimens, and thickness measurements were taken at three locations using an optical system, yielding an average tissue thickness of 0.22 mm. A random speckle pattern was applied to the surface of the tissue using black paint to facilitate displacement tracking via DIC, with fiducial markers added to ensure accurate measurements.

The prepared specimen was mounted on a biaxial testing system (BioTester, CellScale), with an effective testing area of 8.72 $\times$ 10.75 mm. Prior to mechanical testing, the specimen underwent preconditioning in phosphate-buffered saline (PBS) to mimic in vivo conditions, following a protocol of 10 cycles of equi-biaxial tension loading and unloading, aiming for a first Piola-Kirchhoff stress of 150 kPa.

Subsequent biaxial tension tests involved seven different loading protocols, each with different biaxial stress ratios ($P_{11}:P_{22}$), including combinations of 1:1, 1:0.75, 1:0.5, 1:0.25, 0.75:1, 0.5:1, and 0.25:1. Each protocol was repeated for five cycles of loading and unloading, with force and actuator position data recorded at a frequency of 5 Hz, resulting in a total of 1926 data points.

To minimize stress concentration effects near the mounting hooks, deformation analysis focused on the central region of the specimen (4.4 $\times$ 4.4 mm). The force-displacement data were smooth, thanks to the high-resolution load cell with 0.1\% accuracy. The displacement rates used during testing (0.16 mm/s in the X-direction and 0.28 mm/s in the Y-direction) were low enough to avoid significant viscoelastic effects, ensuring the tests were quasi-static with minimal hysteresis. This approach allowed for reliable mechanical characterization of the tissue’s biaxial behavior under the given conditions.

\section{Experimental Details} \label{app:exp}
\subsection{Details in PGS implementation}
In the PGS implementation, we set the maximum number of iterations to 50,000. The initial learning rate is 1e-5, with a learning rate decay factor of 0.9 applied every 1,000 steps. The basis function is a first-order polynomial. The training stopping criteria parameters $\text{TOL}_{\ub}$ and $\text{TOL}_{\Eb}$ are set to 3 and 1e-5,  respectively.

\begin{figure}[h]
    \centering
    \includegraphics[width=0.58\linewidth]{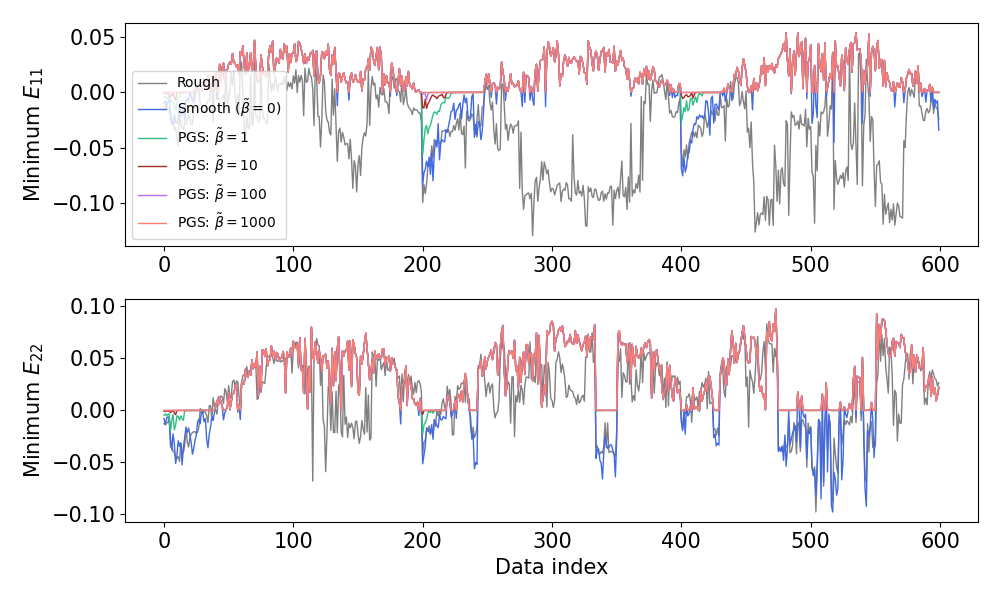}
    \includegraphics[width=0.40\linewidth]{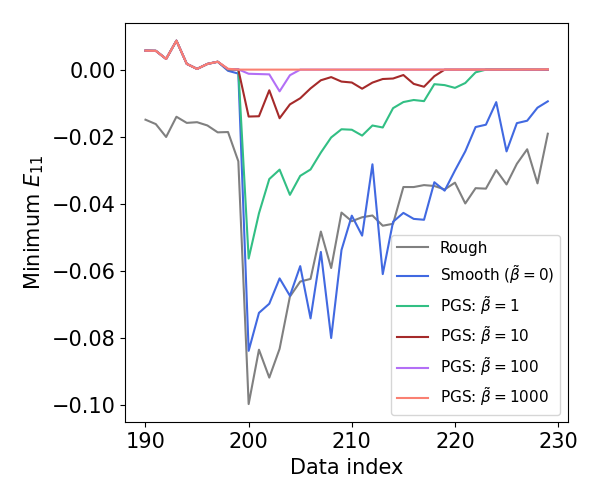}
    \caption{Dataset 1: Comparison of strains of PGS data obtained using different $\tilde{\beta}$.}
    \label{fig:data1_strain_beta} 
\end{figure}



\begin{figure}[h]
    \centering
    \includegraphics[width=1.0\linewidth]{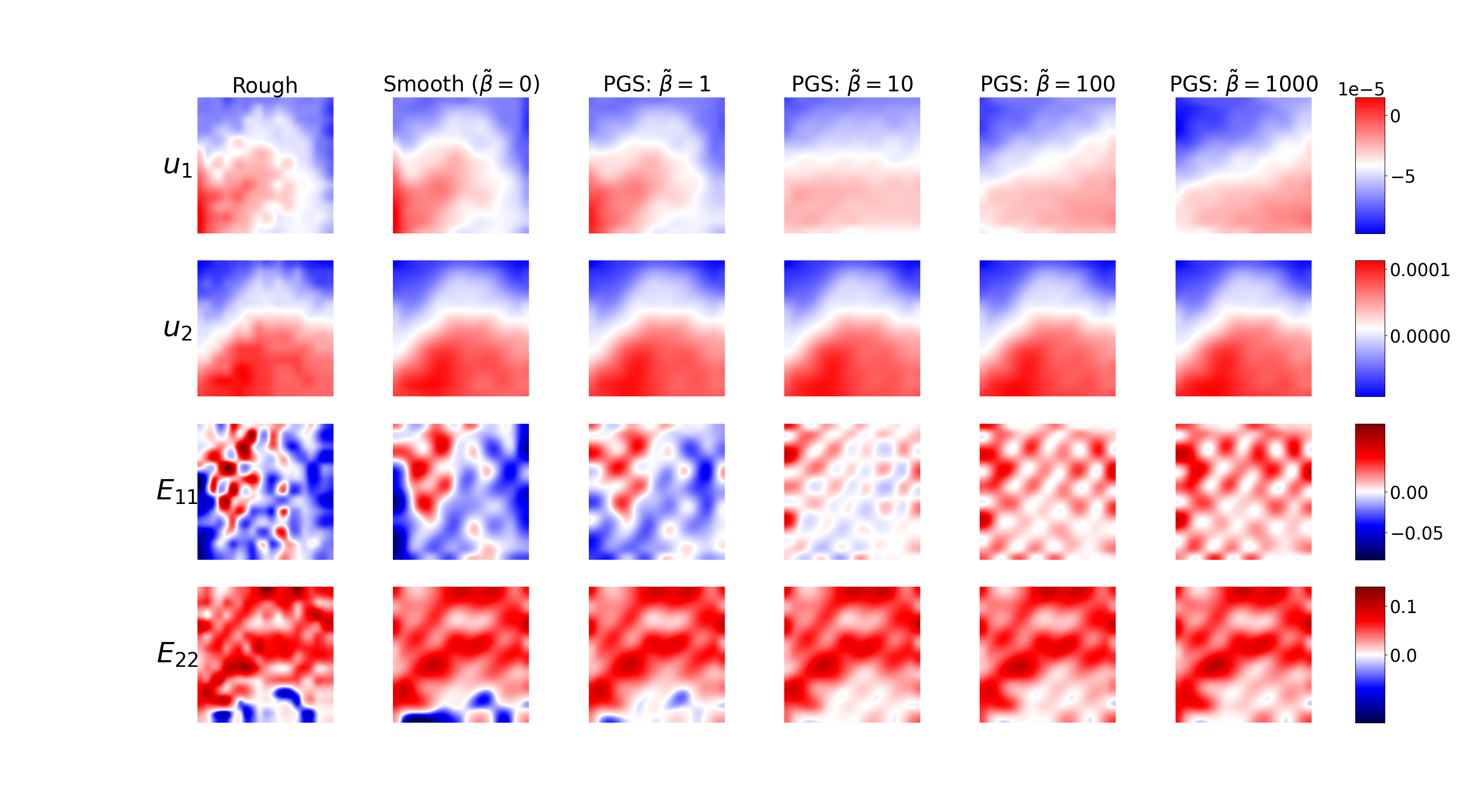}
    \caption{Dataset 2: Comparison of the displacements and strains when using different weight parameters $\tilde{\beta}$.}
    \label{fig:data1_u_strain_beta_1_201} 
\end{figure}

\begin{figure}[h]
    \centering
    \includegraphics[width=0.5\linewidth]{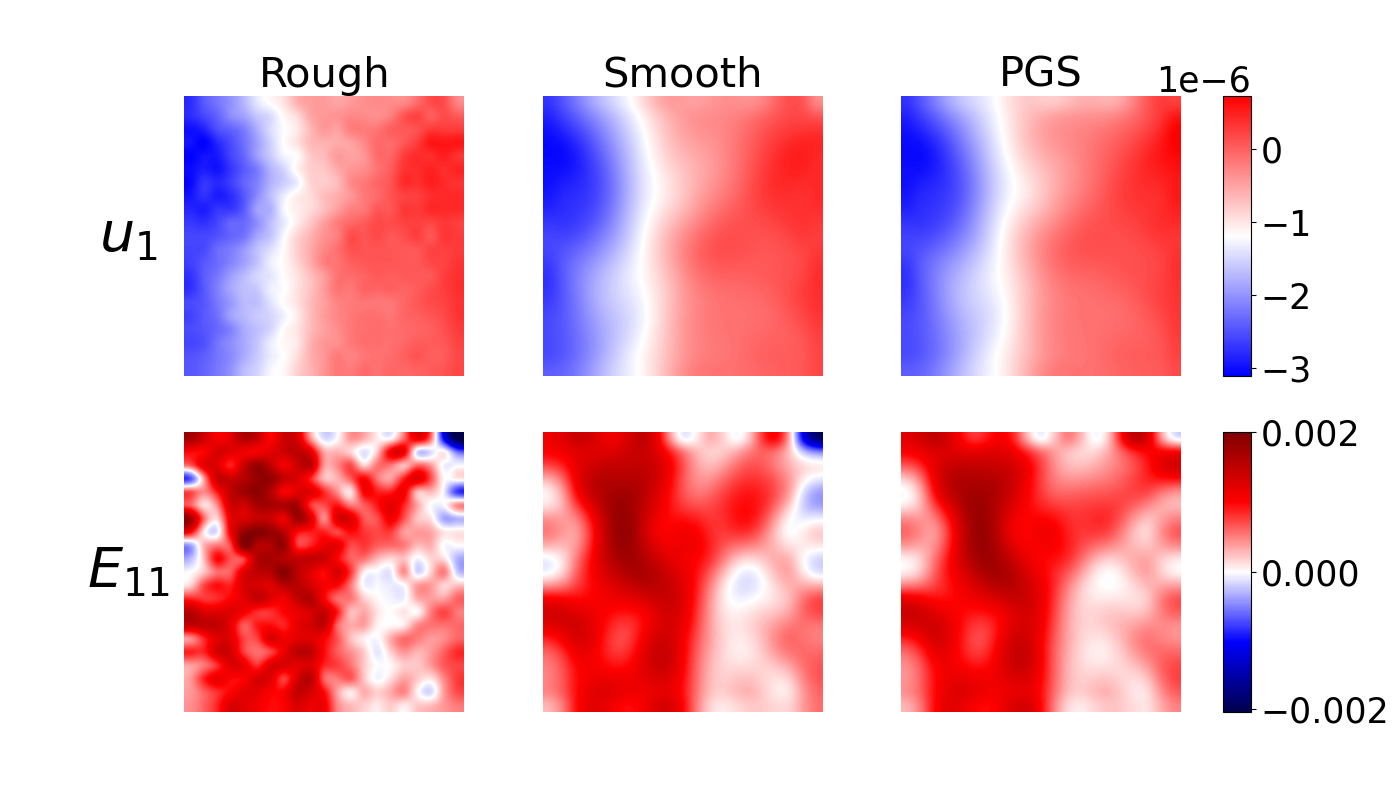}
        \caption{Dataset 2: Comparison of displacement $u_1$ and strain $\epsilon_{11}$ from three datasets.}
    \label{fig:data2_sample} 
\end{figure}

\subsection{Details in PNO training}
 The widths of the MLPs for $\omega^{NN}$, $t^{NN}$ and $\alpha^{NN}$ are (2, 256, 512, 1), (4, 512, 512, 1), and (2, 128, 128, 1) respectively, and the peridynamic horizon size is set as $\delta = 3\Delta x$.  In all experiments we decrease the learning rate with a ratio of decay rate every 50 epochs. To achieve the best training results on each dataset, the initial learning rate is selected from {1e-3, 5e-4, 1e-4}, and the learning rate decay factor is chosen from {0.5, 0.6, 0.8}, with the learning rate decayed every 50 epochs.

\section{Additional Results}\label{app:result}

Herein, we provide more results as a supplement of Section \ref{sec:exp}. 

As a supplement of the ablation study, in Fig. \ref{fig:data1_strain_beta}. we show the minimum strains when using different levels of physical loss parameter, $\tilde{\beta}$. It can be seen that when $\tilde{\beta}$ is too small (e.g., $\tilde{\beta}=1$), the algorithm can not sufficiently eliminates all negative strains. When $\tilde{\beta}$ gets larger, the discrepancy between observed $\ub_{\text{exp}}$ and the reconstructed $\ub$ also increases. From  Fig. \ref{fig:data1_strain_beta} and \ref{fig:data1_u_strain_beta_1_201}, it looks like $\tilde{\beta}=100$ provides a good balance between these two effects.

As a supplement of experiments for dataset 2, in Fig. \ref{fig:data2_sample} we demonstrate the reconstructed strain fields in an exemplar sample, from PGS and the two baselines. One can see that this dataset has smaller negative strain regimes, and PGS still successfully eliminates them without drastically changing the displacement field.

\end{document}